\documentstyle[12pt]{article}
\def\beq{\begin{equation}}
\def\eeq{\end{equation}}
\def\bea{\begin{eqnarray}}
\def\eea{\end{eqnarray}}
\def\nn{\nonumber}
\def\ba{\begin{array}}
\def\ea{\end{array}}
\def\d{\partial}
\def\v{\vert}

\def\r{\rangle}
\def\one{1\hskip -1mm{\rm l}}

\begin{document}
\baselineskip16pt
\smallskip
\begin{center}
{\large \bf \sf
       Algebraic Bethe ansatz for a quantum integrable \\
     derivative nonlinear Schr${\rm {\ddot o}}$dinger model }

\vspace{1.3cm}

{\sf B. Basu-Mallick\footnote{ 
e-mail address: biru@theory.saha.ernet.in }
and Tanaya Bhattacharyya\footnote{E-mail address:
tanaya@theory.saha.ernet.in } },

\bigskip

{\em Theory Group, \\
Saha Institute of Nuclear Physics, \\
1/AF Bidhan Nagar, Kolkata 700 064, India } \\
\bigskip

\end{center}

\vspace {1.1 cm}
\baselineskip=16pt
\noindent {\bf Abstract }

We find that the quantum monodromy matrix associated with a derivative
nonlinear Schr${\rm {\ddot o}}$dinger (DNLS) model exhibits $U(2)$ or 
$U(1,1)$ symmetry depending on the sign of the related coupling constant. 
By using a variant of quantum inverse scattering method which is directly
applicable to field theoretical models, we derive all possible commutation 
relations among the operator valued elements of such monodromy matrix. Thus,
we obtain the commutation relation between creation and annihilation 
operators of quasi-particles associated with DNLS model and find out the 
$S$-matrix for two-body scattering. We also observe that, for some special 
values of the coupling constant, there exists an upper bound on the number 
of quasi-particles which can form a soliton state for the quantum DNLS model.

\baselineskip=16pt
\vspace {.6 cm}

\vspace {.1 cm}

\newpage

\baselineskip=16pt
\noindent \section {Introduction }
\renewcommand{\theequation}{1.{\arabic{equation}}}
\setcounter{equation}{0}

\medskip

Quantum integrable field models and spin chains in low 
dimensions have recently attracted much attention due to their
close connection with diverse areas of physics as well as mathematics. 
By using algebraic Bethe ansatz,
 which occurs naturally in the 
framework of quantum inverse scattering method (QISM),  one can find out
the spectrum and various correlation functions of quantum integrable 
models with short-range interactions [1-4].  
The nonlinear Schr${\rm {\ddot o}}$dinger 
 (NLS) model is a  well known example 
of such quantum integrable field model in $1+1$-dimension [1-8].
 For the case of derivative nonlinear Schr${\rm {\ddot o}}$dinger 
(DNLS) model, however, the situation is a little bit 
complicated due to the following reason.
There exists one type of classical DNLS model with  
 equation of motion like [9]
\beq
iq_t + q_{xx} - 4i \xi (q^*q^2)_x = 0  \, ,
\label {a1}
\eeq
where the subscripts denote the derivatives with respect to
corresponding variables.  By using an
 equal time `nonultralocal' Poisson bracket (PB) structure given by
$\{ q(x), q(y) \} = \{ q^*(x), q^*(y) \} =0, ~
\{ q(x), q^*(y) \} = \delta_x(x-y)$,
  one can show that the infinite 
number of conserved quantities (including the Hamiltonian)
associated with this DNLS equation of motion
yield vanishing PB relations among themselves [10]. This fact establishes
 the classical integrability of DNLS model (\ref {a1}) in the Liouville sense. 
However, due to the appearance of nonultralocal commutation relations
among the basic field operators, quantum version of such DNLS model 
 can not be handled by QISM and therefore quantum integrability 
can not be established for this case. 

On the other hand,
there exists another type of DNLS model with equation of motion like [11]
\beq
i\psi_t + \psi_{xx} - 4i \xi ( {\psi}^*{\psi} ){\psi}_x = 0.
\label {a2}
\eeq
The Hamiltonian 
\beq
 H =  \int_{-\infty}^{+\infty} \big( - \psi^*\psi_{xx} + 2i\xi ( \psi^*\psi )
\psi^*\psi_x \big)dx \, , 
\label {a3}
\eeq 
 and an equal time `ultralocal' PB structure
\beq
\{ \psi (x), \psi(y) \} = \{ \psi^* (x), \psi^*(y) \} = 0, ~~
\{ \psi (x), \psi^*(y) \} = -i \delta(x-y) \,  ,
\label {a4}
\eeq
yield eqn.(\ref {a2}) through
 a canonical evolution: $\psi_t = \{ \psi , H \}$. 
By using the ultralocal PB structure (\ref {a4}) and Lax operator 
 given by 
\bea
         \tilde {U}( x, \lambda )  =    
i \pmatrix { \xi \psi^*(x)\psi(x) -{\lambda^2}/4 & 
\sqrt{\xi}\lambda\psi^*(x) \cr 
\sqrt{\xi}\lambda\psi(x) & - \xi \psi^*(x)\psi(x) + {\lambda^2}/4}  \, ,
\label {a5}
\eea
where $\lambda $ is a spectral parameter,
one can  prove the classical integrability for the DNLS model
(\ref {a2}) in the 
Liouville sense.  Furthermore, by taking advantage of ultralocal commutation 
relations among the basic field operators, integrability of the 
corresponding quantum DNLS model can also be established through QISM [12].

Similar to the case of other integrable systems, the 
monodromy matrix plays a key role in deriving the conserved quantities
and studying exact solvability
  of the DNLS model (\ref {a2}) as well as its quantum analogue. 
  Though the PB relations (commutation relations)
 among various elements of classical (quantum) monodromy matrix 
associated with such DNLS model have been computed earlier [12], 
this  problem should be reinvestigated 
due to the following reasons. First of all, it was assumed 
earlier that the monodromy matrix of classical (quantum) DNLS model
 exhibits  $SU(2)$ ($U(2)$) symmetry 
for any value of the corresponding coupling constant $\xi $.
However, in comparison to the case of 
 NLS model where the symmetry of  monodromy matrix depends on the 
sign of the corresponding coupling constant, this seems to be a rather
questionable assumption. Therefore, it is necessary to properly
study the symmetry properties of the classical (quantum)
 monodromy matrix of DNLS model
and use those symmetries as an input for deriving the PB relations 
(commutation relations) among the elements of the monodromy 
matrix. Secondly, a rather cumbersome method was adopted earlier for 
finding out the commutation relations 
among the elements of quantum monodromy 
matrix at the infinite interval limit. 
  As it will be clear from the discussions in
Sec.4 of the present article, the above mentioned
 method yields an incorrect commutation 
relation between creation and annihilation 
operators associated with the quasi-particles of quantum DNLS model.
Since such commutation relation plays a crucial role 
in finding out the norm of Bethe eigenstates, the $S$-matrix 
for two-body scattering and various correlation 
functions, it is necessary to calculate very carefully
the commutation relations among the elements of quantum monodromy 
matrix at infinite interval limit. Finally, it may be noted that two different 
approaches were taken earlier for treating the classical and quantum 
DNLS model [12]. The integrability of quantum DNLS model was established
by first discretising the system on a lattice, evaluating the 
commutation relations among the elements of the monodromy matrix 
defined on that lattice and finally taking the continuum limit.  
For the case of classical DNLS model, however, no such lattice regularisation 
was taken and PB relations among the elements of monodromy matrix 
were evaluated directly for the continuum model.

The aim of the present article is to shed some light on the above mentioned 
issues and especially study the quantum DNLS model 
by using a variant of QISM [2] which can be applied to 
continuum field models without performing any lattice regularisation.
In Sec.2 of this article,
we find out the symmetry properties 
of monodromy matrix associated with the 
classical DNLS model (\ref {a2}) and subsequently use those symmetries 
for evaluating PB relations among various elements of this
monodromy matrix. In Sec.3, we construct 
 the quantum monodromy matrix of DNLS model 
on a finite interval and derive all possible commutation relations 
among the elements of such monodromy
matrix through QISM.  In Sec.4, we take the 
 infinite interval limit of these commutation relations  and construct
 exact eigenstates for the diagonal elements of quantum monodromy matrix
through algebraic Bethe ansatz. Furthermore,
we obtain the commutation relation between creation and annihilation 
operators of quasi-particles associated with DNLS model
and find out the $S$-matrix of two-body scattering among such
quasi-particles.  Sec.5 is the concluding section.

\vspace{1cm}

\noindent \section {Integrability of the classical DNLS model }
\renewcommand{\theequation}{2.{\arabic{equation}}}
\setcounter{equation}{0}

\medskip

To investigate the symmetry properties 
of monodromy matrix associated with the 
classical DNLS model (\ref {a2}),
 we start with a Lax operator of the form 
\bea
          U( x, \lambda )  =   
i  \pmatrix { \xi \psi^*(x)\psi(x)-{\lambda^2}/4 & 
\xi\lambda\psi^*(x) \cr 
\lambda\psi(x) & - \xi \psi^*(x)\psi(x) + {\lambda^2}/4} \, .
\label{b1}
\eea
Note that the off-diagonal elements of this
Lax operator differ from the corresponding elements of previously given 
Lax operator (\ref {a5}) through some scale factors. However, as will be 
shown later, the traces of monodromy matrices associated 
with the Lax operators (\ref {a5}) and (\ref {b1}) coincide with each other
and consequently both of these Lax operators correspond to the same 
DNLS model (\ref {a2}).

By using the Lax operator (\ref {b1}) along with its asymptotic form at 
$\v x \v \rightarrow \infty $, we define the monodromy matrices 
of DNLS model on finite and infinite intervals as
\beq
T^{x_2}_{x_1}(\lambda) = {\cal P} \exp \int_{x_1}^{x_2} U(x,\lambda) dx 
\label{b2}
\eeq
and 
\beq
T(\lambda) =  
\lim_{\stackrel {x_2 \rightarrow + \infty} {x_1 \rightarrow -\infty}} 
e(-x_2,\lambda) \left \{
{\cal P} \exp \int_{x_1}^{x_2} U(x,\lambda) dx \right \} e(x_1,\lambda) 
\label{b3}
\eeq
respectively, where ${\cal P} $ denotes the path ordering and 
$ e(x,\lambda) 
= e^{-{\frac{i}{4}} \lambda^2 \sigma_3 x} $. To find 
out the symmetries of such monodromy matrices, we note that 
 the Lax operator (\ref {b1}) satisfies the relations
\bea
~~~~~~~~~~~U ( x,\lambda )^* = K \, U ( x,\lambda^* ) \, K^{-1} , ~~~~~~
U ( x, -\lambda ) = L \, U(x,\lambda ) \, L^{-1} \, ,
~~~~~~~~~~~~~~~~~ ( 2.4 a,b ) \nn
\eea
\addtocounter {equation}{1} 
where $ K = \pmatrix { 0 & \sqrt {-\xi} \cr {1 / \sqrt{-\xi}} & 0 } $ and 
 $ L = \pmatrix { 1 & 0 \cr 0 & -1 } $. 
 By using these relations we find 
 that the symmetries of monodromy matrix (\ref{b3}) are given by
\bea
~~~~~~~~~~~~~~~T(\lambda)^* = K \, T( \lambda^* )\, K^{-1}  ,
~~~~~~  T ( -\lambda ) = L \, T( \lambda ) \, L^{-1}
 \, . ~~~~~~~~~~~~~~~~~~~~~~~~~~(2.5 a,b ) \nn
\eea
\addtocounter{equation}{1}
By exploiting the symmetry relation (2.5a) and restricting to the case when
$\lambda $ is a real 
parameter, one may now express $T(\lambda )$ in the form 
\bea
T ( \lambda ) = \pmatrix { a( \lambda ) & -\xi b^*( \lambda ) \cr
                          b( \lambda ) & a^*( \lambda ) } \, .
\label{b6}
\eea 
Since the Lax operator (\ref {b1}) is a traceless matrix, we also get 
 $ det {T( \lambda )} = 1$ or, equivalently,  $ |{a( \lambda )}|^2 +
\xi|{b(\lambda)}|^2 = 1 $.
 Moreover, by using the symmetry relation 
(2.5b), it is easy to see
 that $a(-\lambda)=a(\lambda)$ and $b(-\lambda )=-b(\lambda)$.
So, for the case of real $\lambda $,
$T(\lambda)$ within the range $\lambda\geq 0 $ 
contains all informations 
about the scattering data. Therefore, in the following we shall consider 
the PB relations among the elements of monodromy matrix (\ref {b6})
only within the range $\lambda\geq 0 $. 

In analogy with the monodromy matrix $T(\lambda )$ (\ref {b3})
which is defined through Lax operator (\ref {b1}), 
one can also define the monodromy matrix ${\tilde T}(\lambda)$
corresponding to Lax operator (\ref {a5}) as 
\beq
{\tilde T}(\lambda) =  
\lim_{\stackrel {x_2 \rightarrow + \infty} {x_1 \rightarrow -\infty}} 
e(-x_2,\lambda) \left \{
{\cal P} \exp \int_{x_1}^{x_2} 
{\tilde U}(x,\lambda) dx \right \} e(x_1,\lambda) \, .
\label{b7}
\eeq
It was assumed earlier that this ${\tilde T}(\lambda)$ can be written
in a $SU(2)$ symmetric form 
for any value of the corresponding coupling constant $\xi $ [12].
For the purpose of properly investigating the 
symmetry properties of ${\tilde T}(\lambda)$,  we observe that
 the Lax operators (\ref {a5}) and (\ref {b1}) are related through a 
symmetry transformation given by
\beq
M U( x, \lambda ) M^{-1} = \tilde{U} ( x, \lambda ) \, ,
\label{b8}
\eeq
where $M = \pmatrix { \xi^{-\frac{1}{4}} & 0 \cr 0 & \xi^{\frac{1}{4}} } $.
Consequently, the corresponding monodromy matrices would also be related
 as $ \tilde{T}( \lambda ) = M T( \lambda ) M^{-1} $. By using this relation 
along with eqn.(\ref {b6}), 
we can express $\tilde{T}(\lambda)$ for real $\lambda $ as 
\beq
\tilde{T}( \lambda ) = \pmatrix { \tilde{a}( \lambda
) & -\rho {\tilde{b}}^*( \lambda ) \cr
\tilde{b}( \lambda ) & {\tilde{a}}^*( \lambda )} \, ,
\label{b9}
\eeq
where $\rho = {\rm sign} \, \xi$ and 
\bea
~~~~~~~~~~\tilde{a}(\lambda) = a(\lambda ) , ~~ {\tilde{a}}^*( \lambda ) 
= a^*( \lambda), ~~ \tilde{b}(\lambda)=\sqrt{\xi} \, b(\lambda ), 
~~ {\tilde{b}}^*( \lambda )=\rho \sqrt{\xi} \, b^*( \lambda ) \, .  
\label{b10}
\eea
Now from eqn.(\ref {b9}) it is evident that the monodromy matrix 
$ \tilde{T}( \lambda )$ takes the form of a $SU(2)$ group valued 
object when $\xi >0$ and $SU(1,1)$ group valued object when 
$\xi <0$. Thus we find that, similar to the case of NLS model,
 symmetry of the monodromy matrix associated with DNLS model 
is also determined through the sign of the corresponding coupling constant.  
As a result PB relations among the scattering data of DNLS model,
which were derived earlier 
 by assuming  ${\tilde T}(\lambda)$ to be a 
 $SU(2)$ group valued object,  actually correspond to the case $\xi >0$.

It may be noted that, the 
 diagonal elements of the monodromy matrices $T(\lambda )$
 (\ref {b6}) and $ {\tilde T}( \lambda )$ (\ref {b9}) coincide with 
each other.  So, by using the results of Ref.12 where 
 $\ln {\tilde a}( \lambda ) $ was expanded in powers of ${1 \over \lambda}$,
we can write
\beq
 \ln a( \lambda ) = 
 \ln {\tilde a}( \lambda ) = 
\sum_{n=0}^{\infty} \frac{ i \, C_n}{\lambda^{2n}}  \, ,
\label {b11}
\eeq
and find out the first few $C_n$s as 
\bea
&~~~~~~~~~~~~~~~~~C_0 = -\xi\int_{-\infty}^{+\infty}\psi^*\psi \, dx, ~~~~~~
C_1 =4 i\xi\int_{-\infty}^{+\infty}\psi^*\psi_x \,  dx ,  
~~~~~~~~~~~~~~~~&(2.12a,b) \nn \\
&~~~~~~~~C_2 = 8\xi\int_{-\infty}^{+\infty} \big (
\psi^*\psi_{xx} - 2i\xi ( \psi^*\psi )
\psi^*\psi_x \big ) \, dx . ~~~~~~~~~~&~~(2.12c) \nn
\eea   
\addtocounter{equation}{1}
Note that the Hamiltonian (\ref {a3}) of DNLS model  is related 
to the third expansion coefficient (2.12c)
 as $ H = - \frac{1}{8 \xi} C_2 $. As a result, the monodromy matrices 
   ${\tilde T}(\lambda )$ (\ref {b9}) and $T( \lambda )$ (\ref {b6})
 correspond to the same DNLS model (\ref {a2}). 

Next, we want to derive the PB relations among the elements of 
    $T( \lambda )$ (\ref {b6}) for both positive and 
negative values of the coupling constant $\xi $. To this end,
 we apply (\ref {a4}) to evaluate the PB relations among the elements of 
Lax operator (\ref {b1}) and find that 
\beq
\left\{ U( x, \lambda ) {\stackrel {\otimes}{,}}
 U( y, \mu ) \right\} = \left [ \, r( \lambda, \mu ), U( x,
\lambda ) \otimes \one + \one \otimes U( y, \mu ) \, 
\right ]\, \delta( x-y ) \, ,
\label{b13}
\eeq  
where 
\beq
r( \lambda, \mu ) = -\xi \big ( \, t^c\sigma_3\otimes\sigma_3 + s^c (
\sigma_+\otimes\sigma_- + \sigma_-\otimes\sigma_+ ) \, \big )
\label{b14}
\eeq
with $ t^c = \frac{\lambda^2 + \mu^2}{2( \lambda^2 - \mu^2 )} , \quad s^c =
\frac{2\lambda\mu}{\lambda^2 - \mu^2} $. Next, by employing
a standard technique [1] for deriving PB relations among the elements of 
 monodromy matrix (\ref {b3}) with the help of eqn.(\ref {b13}),
we obtain
\beq
\left\{ T( \lambda ){\stackrel{\otimes}{,}} T( \mu ) \right\} 
= r_+ ( \lambda, \mu ) T( \lambda )
\otimes T( \mu ) - T( \lambda ) \otimes T ( \mu )r_-( \lambda, \mu ) \, ,
\label{b15}
\eeq
where $r_\pm (\lambda , \mu) $ matrices are given by
\bea
r_\pm (\lambda , \mu )
&=& E^{\pm1}( -L, \lambda ) \otimes E^{\pm1}( -L, \mu )r( \lambda,\mu )
E^{\mp1}( -L, \lambda ) \otimes E^{\mp1}( -L, \mu )  \nn \\
&=& -\xi \left ( t^c \sigma_3 \otimes \sigma_3 + s^c_\pm \sigma_+ 
\otimes \sigma_- + s^c_\mp \sigma_- \otimes \sigma_+ \right ) \, ,
\nn
\eea 
with $ s^c_\pm = \pm 2 i \pi \lambda^2 \delta ( \lambda^2 - \mu^2 ) $. 
By substituting the symmetric form of $T(\lambda )$  (\ref {b6}) 
to eqn.(\ref{b15}) and expressing it
 in elementwise form, we finally obtain 
\bea
&&~~~~~\{ a( \lambda ) , a( \mu ) \} = 0 \, , ~~
\{ a( \lambda ) , a^\dagger( \mu ) \} = 0 \, , 
~~~~~~~~~~~~~~~~~~~~~~~~~~~~~~~~~~~~~~~~~~~~~ \nn (2.16a,b) \\
&&~~~~~\{ a( \lambda ) , b( \mu ) \} = \xi \left(
\frac{\lambda^2 + \mu^2 }{\lambda^2 - \mu^2} \right) \,
a( \lambda )b( \mu ) - 2i\pi\xi\lambda^2 \, \delta( \lambda^2 - \mu^2 ) \, 
b(\lambda )a(\mu)  \, ,\nn ~~~~~~~~~( 2.16 c ) \\
&&~~~~~\{ a( \lambda ) , b^*( \mu ) \} = -\xi \left(
\frac{\lambda^2 + \mu^2 }{\lambda^2 - \mu^2} \right) \,
a( \lambda )b^*( \mu ) + 2i\pi\xi\lambda^2 \, \delta( \lambda^2 - \mu^2) \,
b^*(\lambda )a( \mu ) \, , \nn ~~~(2.16 d) \\
&&~~~~~\{ b( \lambda ) , b^*( \mu ) \} 
= -4i \pi \lambda^2 \, \delta( \lambda^2 -
\mu^2) \, \v {a( \lambda )} \v^2 \, . \nn ~~~~~~~~~~~~~~~~~~~~~~~~~~~~
~~~~~~~~~~~~~~(2.16e)
\eea
\addtocounter{equation}{1}
The above PB relations among the scattering data of the DNLS model
 are evidently valid for all
values of the coupling constant $\xi $.
Since from eqns.(2.16a) and (\ref {b11}) it follows
that $\{C_m,C_n\}=0$ for all $m,n$, 
 the DNLS model (\ref {a2}) represents a classical integrable
system in the Liouville sense. 
With the help of 
  transformation (\ref {b10}), one can also recast eqns.(2.16a-e) as the  
 PB relations among the elements of monodromy matrix 
   ${\tilde T}(\lambda )$ (\ref {b9}) 
and compare such PB relations with their counterparts in Ref.12.  It is 
easy to see that the forms of eqns.(2.16a-d) remain unaltered 
through the transformation (\ref {b10}) and 
coincide with their counterparts as given earlier
(after taking care of slight changes in notations and in
 the definition of fundamental PB relation). 
However, by using
eqns.(2.16e) and (\ref {b10}) we find that
\beq
\{ \tilde{b}( \lambda ) , \tilde{b^*}( \mu ) \} = -4i\pi \v \xi
\v \lambda^2 \delta( \lambda^2 - \mu^2 ) \v {{\tilde a}( \lambda )} \v^2.
\label{b17}
\eeq
It is interesting to note that, for the case $\xi < 0$,
  the above equation differs
 from its counterpart [12] through a sign factor.  
Thus, we are able to derive here 
the correct PB relation between 
$\tilde{b}( \lambda ) $ and  $\tilde{b^*}( \mu )$ for 
the case $\xi < 0$. It is well 
known that the commutation relation between the quantum analogues of 
$\tilde{b}( \lambda ) $ and  $\tilde{b^*}( \mu )$ plays a crucial
role in finding out the norm of the Bethe eigenstates, the $S$-matrix 
for two-body scattering and various correlation functions.  
As a first step for 
   evaluating this commutation relation and other
commutation relations which would represent the 
 quantum counterparts of the classical PB relations
(2.16), in the following we shall quantise
 the monodromy matrix (\ref {b2}) of DNLS model on a finite interval. 

\vspace{1cm}

\noindent \section 
{Commutation relations for the quantum monodromy matrix on a finite interval}
\renewcommand{\theequation}{3.{\arabic{equation}}}
\setcounter{equation}{0}

\medskip

In the quantised version of the DNLS model (\ref{a2}), the fundamental PB 
relations (\ref {a4}) are replaced by 
equal time commutation relations among the basic
field operators:
\beq
\left[ \psi( x ) , \psi( y ) \right] = 
\left[\psi^\dagger( x ) ,\psi^\dagger( y )
\right] = 0 , ~~~~
\left[ \psi( x ) ,\psi^\dagger( y ) \right] =  \delta( x - y ) \, ,
\label {c1}
\eeq
and vacuum state is defined as $\psi (x) \v 0 \r = 0$.
By using the ultralocal commutation relations (\ref {c1}) and a version 
of QISM which can be applied to field models without performing any lattice
regularisation [2], at present we shall
construct the quantum monodromy matrix of DNLS model
on a finite interval and 
 derive all commutation relations among the elements 
of such quantum monodromy matrix.
 To this end, we assume that the quantised form of the
classical Lax operator (\ref {b1}) is given by
\bea
{\cal U}_q( x, \lambda ) = i \pmatrix { f\rho( x ) - \lambda^2 /4 &
\xi\lambda\psi^\dagger( x )  \cr
\lambda\psi(x)  &  -g\rho( x ) + \lambda^2/4 }
\label{c2}
\eea
where $\rho( x ) = \psi^\dagger( x )\psi( x )$, and $f, \, g$ are 
two yet undetermined parameters. It will be shown later that 
$f \rightarrow \xi $,
$g \rightarrow \xi $ at the classical limit and, as a result,
${\cal U}_q( x, \lambda )$ (\ref {c2}) correctly reproduces
$ U( x, \lambda )$ (\ref {b1}) in this limit.
By using ${\cal U}_q( x, \lambda )$,  we define 
the quantum monodromy matrix of DNLS model on a finite interval as 
\beq
{\cal T} ^{x_2}_{x_1}(\lambda) = \quad  :{\cal P} \exp \int_{x_1}^{x_2} {\cal
U}_q(x,\lambda) dx : \, ,
\label {c3}
\eeq
where the symbol $::$ denotes the normal ordering of operators. It is 
evident that this quantum monodromy matrix 
(\ref {c3}) satisfies 
a differential equation given by
\bea
\frac{\partial}{\partial x_2}{\cal T}^{x_2}_{x_1}( \lambda ) &=&
 : {\cal U}_q( x_2 ,
\lambda ){\cal T}^{x_2}_{x_1}( \lambda ) : \nn\\
\quad \quad \quad 
&=& - \frac{i}{4}\lambda^2\sigma_3{\cal T}^{x_2}_{x_1}( \lambda ) +
 i\xi\lambda\psi^\dagger(x_2)\sigma_+{\cal T}^{x_2}_{x_1}( \lambda ) +
 i\lambda\sigma_-{\cal T}^{x_2}_{x_1}( \lambda )\psi( x_2 )\nn\\
&~&~~+if\psi^\dagger( x_2 )e_{11}{\cal T}^{x_2}_{x_1}( \lambda )\psi( x_2 )
- ig\psi^\dagger( x_2 )e_{22}{\cal T}^{x_2}_{x_1}( \lambda )\psi( x_2 ) \, ,
\label{c4}
\eea  
where $e_{11} = \frac{1}{2}( 1 + \sigma_3 )$ and 
$e_{22} = \frac{1}{2}(1 - \sigma_3 ) $. 
For the purpose of applying QISM, however,
it is needed to find out the differential equation satisfied by the product 
${\cal T}^{x_2}_{x_1}(\lambda)\otimes {\cal T}^{x_2}_{x_1}(\mu)$.
 To this end, we borrow a notation of Ref.2
where the sign for normal arrangement of operator factors is taken as
$\vdots\vdots $ . The sign $\vdots\vdots$ , applied to the product of
several operator factors (including $\psi$ and $\psi^\dagger$), ensures
the arrangement of all $\psi^\dagger$ on the left, and all $\psi$ on the
right, {\it without altering the order
 of the remaining factors}.  For example,
$$ \vdots X \psi \psi^\dagger Y \vdots = \psi^\dagger X Y \psi \, ,$$
where $X$ and $Y$ may be taken as some elements of the
quantum monodromy matrix (\ref {c3}).
Now by using the basic commutation relations 
(\ref {c1}) and the method of `extension' [2],
we find that the product of two monodromy matrices 
satisfies the following differential equation (details of this 
calculation are given in the Appendix): 
\beq
\frac{\partial}{\partial x_2} 
\left({\cal T}^{x_2}_{x_1}(\lambda)\otimes {\cal T}^{x_2}_{x_1}(\mu) \right) 
= \vdots {\cal L}( x_2; \lambda, \mu )
{\cal T}^{x_2}_{x_1}( \lambda ) \otimes {\cal T}^{x_2}_{x_1}( \mu ) \vdots ~,
\label{c5}
\eeq
where
\beq
{\cal L}( x; \lambda, \mu ) 
= {\cal U}_q( x, \lambda ) \otimes \one + \one \otimes
{\cal U}_q( x, \mu ) + {\cal L}_\triangle(x ; \lambda, \mu ) \, ,
\label{c6}
\eeq
with
$$
{\cal L}_\triangle(x ; \lambda, \mu ) = \pmatrix {-f^2\rho( x ) & 
- \mu \xi f\psi^\dagger( x ) & 0 & 0 \cr
0 & gf\rho( x ) & 0 & 0 \cr
-\lambda f\psi( x ) & - \lambda\mu \xi & gf\rho( x ) & 
\mu \xi g\psi^\dagger( x ) \cr
0 & \lambda g\psi( x ) & 0 & -g^2  \rho( x )} \, .
$$
Next, let us consider a $(4\times 4)$ $R(\lambda, \mu)$ matrix of the form 
\beq
R( \lambda, \mu ) = \pmatrix { 1 & 0 & 0 & 0 \cr
0 & s( \lambda, \mu ) & t( \lambda, \mu ) & 0 \cr
0 & t( \lambda, \mu ) & s( \lambda, \mu ) & 0 \cr
0 & 0 & 0 & 1} \, ,
\label{c7}
\eeq
with $t(\lambda, \mu)=\frac{\lambda^2 - \mu^2}{\lambda^2 q - \mu^2
q^{-1}} , ~ s( \lambda, \mu ) = \frac{( q - q^{-1} )\lambda\mu}
{\lambda^2 q - \mu^2 q^{-1}} $ and $ q = e^{-i\alpha} $,  $\alpha$ being an
yet undetermined real parameter. It is easy to check that the 
 $R(\lambda, \mu)$ matrix (\ref {c7}) and 
${\cal L}( x; \lambda, \mu )$ (\ref {c6}) satisfy an equation given by
\beq
R( \lambda, \mu ){\cal L}( x; \lambda, \mu ) = 
{\cal L}( x; \mu, \lambda )R(\lambda, \mu) \, ,
\label{c8}
\eeq
provided  the parameters $f$, $g$,  $\alpha $ and the coupling
 constant $\xi$ are related as
\bea
~~~~~~~~~~~~~~~~~~~~~~~\xi= -\sin\alpha , 
~~~~ f = \frac{\xi e^{-i\alpha / 2}}{\cos \alpha / 2},~~~~
g = \frac{\xi e^{i\alpha / 2}}{\cos \alpha / 2} \, . \nn
~~~~~~~~~~~~~~~~~~~~~~  (3.9a,b,c)
\eea
\addtocounter{equation}{1}
By using eqns.(\ref{c5}) and (\ref{c8}), we find that
 the monodromy matrix (\ref {c3}) of DNLS model satisfies 
the quantum Yang-Baxter equation (QYBE):
\beq
R( \lambda, \mu ){\cal T}^{x_2}_{x_1}( \lambda ) \otimes 
{\cal T}^{x_2}_{x_1}( \mu ) = {\cal T}^{x_2}_{x_1}( \mu ) \otimes 
{\cal T}^{x_2}_{x_1}( \lambda )R( \lambda, \mu ) \, .
\label{c10}
\eeq
Expressing this QYBE in elementwise form, one may explicitly obtain 
all possible commutation relations among the elements of quantum 
monodromy matrix (\ref {c3}). 

Note that the relations (3.9a,b,c) are obtained as a necessary condition 
 for satisfying QYBE (\ref {c10}).  By solving eqns.(3.9a,b,c),  
the parameters 
$f$, $g$ and $\alpha $ can be determined as some functions of the known
coupling constant $\xi $. 
Thus QYBE fixes 
all undetermined parameters in the quantum Lax operator (\ref {c2}) and 
  corresponding $R(\lambda, \mu)$ matrix (\ref {c7}). 
Due to eqn.(3.9a) it is evident that, 
the coupling constant 
 of quantum  DNLS model must be restricted 
 within a range  given by $\v \xi \v \leq 1$ and the parameter $\alpha $
 has a one-to-one correspondence with such coupling constant for
$-{\pi \over 2} \leq \alpha \leq {\pi \over 2}$.
It may be noticed that, by putting $x_2
= x_1 + \Delta$ in eqn.(\ref {c3})  (here $\Delta $ 
is a small positive parameter) and retaining terms only up to the 
  order $\Delta$, one obtains
\beq
{\cal T} ^{x_1+\Delta }_{x_1}(\lambda) 
\sim \one + i \Delta \pmatrix { f\rho_n - \lambda^2 /4 &
\xi\lambda\psi^\dagger_n  \cr
\lambda\psi_n  &  -g\rho_n + \lambda^2/4 } \, ,
\label{c11}
\eeq
where 
$ \psi^\dagger_n  = {1\over \Delta } \int_{x_1}^{x_1+\Delta} 
\psi^\dagger (x) dx$,
$ \psi_n  = {1\over \Delta } \int_{x_1}^{x_1+\Delta} 
\psi (x) dx$,
$ \rho_n  = {1\over \Delta } \int_{x_1}^{x_1+\Delta} \rho (x) dx$.
It is interesting to observe that this 
${\cal T} ^{x_1+\Delta }_{x_1}(\lambda) $ (\ref {c11})  
satisfies QYBE (\ref {c10}) up to order $\Delta$ and 
reproduces (up to a gauge transformation) the lattice Lax operator 
for quantum DNLS model [12] when the parameter $\Delta $ is identified 
as a lattice constant. In this article, however, we shall not use 
any lattice discretisation and directly work with the 
quantum monodromy matrix (\ref {c3}) which satisfies QYBE exactly.

Next, for the purpose of investigating
 the classical limit of the quantum Lax operator 
(\ref {c2}),  we replace the last commutation relation 
in eqn.(\ref {c1}) by 
$[ \psi( x ) ,\psi^\dagger( y )] = h \delta( x - y )$,
where $h$ is the Planck's constant, and then 
repeat all calculations of this section. It turns out that eqns.(3.9b,c)
remain unchanged, but eqn.(3.9a) is modified as: $h \xi= -\sin\alpha $.
Consequently, for any fixed value of $\xi $, 
$\alpha \rightarrow 0$ limit is essentially equivalent to
 $h\rightarrow 0$ limit. 
Since from eqns.(3.9b,c) it follows that
$f\rightarrow \xi$ and $g\rightarrow \xi$ when 
$\alpha \rightarrow 0$, the quantum Lax operator (\ref {c2})
indeed reproduces the classical Lax operator (\ref {b1})
at $h \rightarrow 0$ limit.

\vspace{1cm}

\noindent \section {Algebraic Bethe ansatz for the quantum monodromy
matrix on an infinite interval}
\renewcommand{\theequation}{4.{\arabic{equation}}}
\setcounter{equation}{0}

\medskip

For the purpose of taking  infinite interval limit 
of QYBE (\ref {c10}), we define the quantum analogue of 
classical monodromy matrix (\ref {b3}) as 
\beq
{\cal T}(\lambda) = 
\lim_{\stackrel {x_2 \rightarrow + \infty} {x_1 \rightarrow -\infty}} 
e(-x_2,\lambda) {\cal T}^{x_2}_{x_1}(\lambda)e(x_1,\lambda) \, ,
\label {d1}
\eeq
where ${\cal T}^{x_2}_{x_1}(\lambda)$ is given by eqn.(\ref {c3}).
 It may be observed that, exactly like the case of classical 
Lax operator (\ref {b1}),
 the quantum Lax operator (\ref {c2}) also satisfies the following relations: 
\bea
~~~~~~~~~~~~~~~
{\cal U}_q( x,\lambda )^* = K  \, {\cal U}_q( x,\lambda^* ) \, K^{-1} ,
\quad  {\cal U}_q( x, -\lambda ) =  L  \,
{\cal U}_q(x,\lambda ) \,  L^{-1}
\, , \nn~~~~~~~~~~~~~~~(4.2a,b)
\eea
\addtocounter{equation}{1}
where $K$ and $L$ are same matrices which have appeared in (2.4a,b).  
By using eqn.(4.2a) and assuming $\lambda $ to be a real parameter,
 is easy to show that the quantum monodromy
matrix (\ref {d1}) can be expressed in a symmetric form given by
\bea
{\cal T}(\lambda)=\pmatrix {A(\lambda) & -\xi B^\dagger(\lambda) \cr
                          B(\lambda) & A^\dagger(\lambda)} \, .
\label {d3}
\eea
Since from eqn.(4.2b) it follows that $A(-\lambda)= A(\lambda)$ and 
$B(-\lambda)= - B(\lambda)$, it is necessary to consider the quantum
monodromy matrix (\ref {d3}) only within the range $\lambda \geq 0$.
In analogy with  ${\cal T}(\lambda )$ (\ref {d1}),
which is defined by quantising the 
 classical monodromy matrix $T(\lambda)$(\ref {b3}),
one may also define the quantum analogue of the classical monodromy matrix 
${\tilde T}(\lambda ) $ (\ref {b7}).
By using arguments similar to the classical 
case, it is easy to show that such quantum monodromy matrix 
can be written in a symmetric form given by
\bea
{\tilde {\cal T}}(\lambda)=\pmatrix {{\tilde A}(\lambda) 
& -\rho {\tilde B}^\dagger(\lambda) \cr
                          {\tilde B}(\lambda) & 
{\tilde A}^\dagger(\lambda)} \, ,
\label {d4}
\eea
where $\rho = {\rm sign} \,  \xi$ and 
${\tilde A}(\lambda) = A(\lambda) , ~ {\tilde B}(\lambda) = \sqrt \xi
B(\lambda) $. 
Thus we find that, the quantum monodromy matrix 
$ {\tilde{\cal T}}( \lambda )$ can be expressed like a $U(2)$ group valued 
object when $\xi >0$ and $U(1,1)$ group valued object when 
$\xi <0$.  As a result, the commutation relations which were derived 
earlier by assuming ${\tilde{\cal T}}(\lambda)$ to be a 
 $U(2)$ group valued object [12] should correspond to the case $\xi >0$.

For finding out the commutation relations among the operator elements
of monodromy matrix (\ref {d1}),
  it is required to get rid of the oscillatory terms which 
arise from the product 
${\cal T}^{x_2}_{x_1}( \lambda ) \otimes {\cal T}^{x_2}_{x_1}( \mu ) $
at the  asymptotic limits $ x_1 , x_2 \rightarrow \pm\infty$.
To this end, we split the 
${\cal L}( x ; \lambda, \mu ) $ matrix (\ref {c6}) into two parts:
$$
{\cal L}( x ; \lambda, \mu ) = {\cal L}_0( \lambda, \mu ) + {\cal L}_1( x;
\lambda, \mu ) \, ,
$$
where $ {\cal L}_0( \lambda, \mu ) $ is given by
$$
{\cal L}_0( \lambda, \mu ) = \lim_{|x| \rightarrow \infty} {\cal L}( x;
\lambda, \mu ) = \pmatrix {-\frac{i}{4}( \lambda^2 + \mu^2 ) & 0 & 0 & 0 \cr
0 & -\frac{i}{4}( \lambda^2 - \mu^2 ) & 0 & 0 \cr
0 & -\xi  \lambda \mu & \frac{i}{4}( \lambda^2 - \mu^2 ) & 0 \cr
0 & 0 & 0 & \frac{i}{4}( \lambda^2 + \mu^2 )} \, ,
$$
and 
$ {\cal L}_1( x ; \lambda, \mu ) $  is the field dependent part of
$ {\cal L}( x; \lambda, \mu ) $, which 
vanishes at $x\rightarrow \pm \infty$.
Due to eqn.(\ref{c8}) it follows that
\beq
R( \lambda, \mu ) \varepsilon( x ; \lambda, \mu ) = \varepsilon( x ; \mu,
\lambda ) R( \lambda, \mu ) \, ,
\label{d5}
\eeq
where $ \varepsilon( x; \lambda, \mu ) = e^{{\cal L}_0( \lambda, \mu )x}$.
By using the above mentioned splitting of   ${\cal L}( x ; \lambda, \mu ) $,
one can derive the integral form of differential equation (\ref{c5}) as
$$
{\cal T}^{x_2}_{x_1}( \lambda ) \otimes {\cal T}^{x_2}_{x_1}( \mu ) = 
\varepsilon( x_2 - x_1; \lambda, \mu ) + \int_{x_1}^{x_2} dx \,
\varepsilon( x_2 - x; \lambda, \mu ) \, \vdots {\cal L}_1( x ; \lambda,\mu )
{\cal T}^{x}_{x_1}( \lambda ) \otimes {\cal T}^{x}_{x_1}( \mu ) \vdots
\, .
$$
Due to the presence of field dependent matrix 
${\cal L}_1(x;\lambda,\mu)$, the second term in the r.h.s. of above 
integral equation vanishes at  
 the limit $ x_1, x_2 \rightarrow \pm\infty $. Consequently, at this
limit,  one gets 
 $\quad{\cal T}^{x_2}_{x_1}( \lambda ) \otimes {\cal T}^{x_2}_{x_1}( \mu )
\rightarrow \varepsilon( x_2 - x_1; \lambda, \mu )$ , which is an
oscillatory term. To get rid of this unwanted term,  we
define an operator like 
\beq
W( \lambda, \mu ) = \lim_{\stackrel
{x_1 \rightarrow -\infty}{x_2 \rightarrow +\infty}}
\varepsilon( - x_2; \lambda, \mu ) 
{\cal T}^{x_2}_{x_1}( \lambda ) \otimes {\cal T}^{x_2}_{x_1}( \mu )
\varepsilon( x_1; \lambda, \mu ) \, ,
\label{d6}
\eeq
which is clearly well behaved at the infinite interval limit.
By using (\ref{c10}) and (\ref{d5}), it is easy to verify that 
this $W( \lambda, \mu )$ (\ref {d6}) satisfies an equation given by
\beq
R( \lambda, \mu ) W( \lambda, \mu ) = W( \mu, \lambda ) R( \lambda, \mu ) \, ,
\label{d7}
\eeq
which represents QYBE for the infinite interval limit.

Next, we want to express QYBE (\ref {d7}) through the direct 
product of two monodromy matrices of the form (\ref {d1}). To this end, 
we note that $W( \lambda, \mu )$ (\ref {d6}) may be rewritten as 
\beq
W( \lambda, \mu ) 
= C_+( \lambda, \mu ) {\cal T}( \lambda ) \otimes {\cal T}(\mu)
 C_-( \lambda, \mu ) \, ,
\label{d8}
\eeq
where
\bea 
C_+( \lambda, \mu ) = \lim_{x \rightarrow \infty} \varepsilon( -x ; \lambda,
\mu ) E( x; \lambda, \mu ) , ~ C_-( \lambda, \mu ) =  \lim_{x \rightarrow
-\infty} E( -x; \lambda, \mu ) \varepsilon( x; \lambda,\mu ) \, , \nn &
~~ (4.9a,b)
\eea
\addtocounter{equation}{1}
with $ E( x; \lambda, \mu ) = e( x, \lambda )\otimes e( x, \mu ) $.
Substituting the explicit  forms of $ E( x; \lambda, \mu ) $  and 
$\varepsilon(x; \lambda,\mu )$ to (4.9a,b), and extracting the limits 
in the principal value sense: 
$\lim_{x \rightarrow \pm \infty}  P( \frac{e^{ikx}}{k} ) $  
$ = \pm i \pi \delta(k)$,  
we obtain 
\bea
C_+( \lambda, \mu ) = \pmatrix { 1 & 0 & 0 & 0 \cr
0 & 1 & 0 & 0 \cr
0 & \rho_+( \lambda, \mu ) & 1 & 0 \cr
0 & 0 & 0 & 1} \, ~~
C_-( \lambda, \mu ) = \pmatrix { 1 & 0 & 0 & 0 \cr
0 & 1 & 0 & 0 \cr
0 & \rho_-( \lambda, \mu ) & 1 & 0 \cr                                 
0 & 0 & 0 & 1}  \, ,
\label {d10}
\eea
where
$$
\rho_\pm( \lambda, \mu ) = \mp \frac{2i\xi  \lambda \mu}{\lambda^2 - \mu^2}
+ 2 \pi \xi  \lambda \mu \delta( \lambda^2 - \mu^2 ) = \mp \frac{2 i \xi 
\lambda \mu}{\lambda^2 - \mu^2 \mp i\epsilon} \, .
$$
By using the expression  (\ref{d8}),
 we rewrite QYBE (\ref {d7}) for the infinite interval limit as 
\beq
R( \lambda, \mu ) C_+( \lambda, \mu ) {\cal T}( \lambda ) 
\otimes {\cal T}( \mu ) C_-( \lambda, \mu ) = 
C_+( \mu, \lambda ) {\cal T}( \mu ) \otimes {\cal T}(\lambda)
 C_-( \mu, \lambda ) R( \lambda, \mu ) \, .
\label{d11}
\eeq 

By inserting the explicit forms of $R(\lambda , \mu)$ (\ref {c7}),
 $C_\pm (\lambda , \mu )$  (\ref{d10}), and 
${\cal T}( \lambda )$ (\ref {d3}) to QYBE (\ref {d11}) and comparing 
its matrix elements from both sides, we finally obtain
\bea
&&\left[ A( \lambda ), A( \mu ) \right] = 0 \, , ~~~~
\left[ A( \lambda ), A^\dagger( \mu ) \right] = 0 \, , \nn 
~~~~~~~~~~~~~~~~~~~~~~~~~~~~~~~~~~~~~~~~~~~~~~~(4.12a,b) \\
&&A(\lambda) B^\dagger(\mu) = \frac{\mu^2 q - \lambda^2 q^{-1}}{\mu^2 -
\lambda^2 - i \epsilon} B^\dagger( \mu ) A( \lambda ) \nn \\
&&~~~~~~~~~~~~~~=  \frac{\mu^2 q - \lambda^2 q^{-1}}{\mu^2 -\lambda^2} 
B^\dagger( \mu ) A( \lambda ) - 2 \pi \xi \lambda \mu \delta( \lambda^2 -
\mu^2 ) B^\dagger( \lambda ) A( \mu ) \, , \nn 
~~~~~~~~~~~(4.12c) \\
&&B( \mu ) A( \lambda ) = \frac{\mu^2 q - \lambda^2 q^{-1}}{\mu^2 -\lambda^2 
- i \epsilon } A( \lambda ) B( \mu ) \nn \\
&&~~~~~~~~~~~~~
=  \frac{\mu^2 q - \lambda^2 q^{-1}}{\mu^2 -\lambda^2} A( \lambda ) B( \mu )
- 2 \pi \xi \lambda \mu \delta( \lambda^2 - \mu^2 ) A( \mu ) B( \lambda)
\, , \nn ~~~~~~~~~~~~~ (4.12d) \\
&&B( \mu ) B^\dagger( \lambda ) = \tau(\lambda,\mu) 
 B^\dagger( \lambda ) B( \mu ) + 4 \pi  \lambda \mu \delta(
\lambda^2 - \mu^2 ) A^\dagger( \lambda ) A( \lambda ) \, , \nn 
~~~~~~~~~~~~~~~~~~~  (4.12e)
\eea
\addtocounter{equation}{1}  
where $q = e^{-i\alpha}$ and 
$\tau( \lambda, \mu ) = \left[ 1 + \frac{8 \xi^2  \lambda^2
\mu^2}{{( \lambda^2 - \mu^2 )}^2} - \frac{4 \xi^2  \lambda^2
\mu^2}{( \lambda^2 - \mu^2 - i \epsilon ) ( \lambda^2 - \mu^2 + i \epsilon
)} \right]$. 
The above commutation relations among the elements of quantum
monodromy matrix  (\ref {d1}) 
 are evidently valid for both positive and negative 
values of the coupling constant $\xi $.
With  the help of  transformations like 
${\tilde A}(\lambda) = A(\lambda) , ~ {\tilde B}(\lambda) = \sqrt \xi
B(\lambda) $, one can also  recast  eqns.(4.12a-e) as 
 the commutation relations among the 
elements of the monodromy matrix ${\tilde {\cal T}}(\lambda )$ (\ref {d4})
and compare such commutation relations with their counterparts in Ref.12. 
  It is easy to see that the forms of commutation relations
(4.12a-d) remain unaltered 
through the above mentioned transformation and  match 
 with their counterparts as given earlier.
However, by using the transformation  
${\tilde A}(\lambda) = A(\lambda) , ~ {\tilde B}(\lambda) = \sqrt \xi
B(\lambda) $, equation (4.12e) may be expressed as 
$$
{\tilde B}( \mu ) {\tilde B}^\dagger( \lambda ) = \tau(\lambda,\mu) 
 {\tilde B}^\dagger( \lambda ) {\tilde B}( \mu )+ 4 \pi \lambda \mu \v \xi \v
\delta(\lambda^2 - \mu^2 ) 
{\tilde A}^\dagger( \lambda ) {\tilde A}( \lambda ) \, , 
$$
which does not match 
at all with its counterpart [12] for either positive or negative
value of $\xi$.  It can be shown that, 
due to a problem which arises while taking 
the infinite interval limit of QYBE,  an incorrect
  commutation relation was obtained earlier between the operators
$ {\tilde B}(\lambda) $ and ${\tilde B}^\dagger (\mu) $ for both 
positive and negative values of $\xi$.
It is interesting to observe that, for the case
$\lambda \neq \mu$, eqn.(4.12e) gives
$[ {\tilde B}(\lambda) , {\tilde B}^\dagger (\mu) ] \neq 0$.
On the other hand, from eqn.(2.16e) it follows that
$ \{ b( \lambda ) , b^*( \mu ) \} =0$ for 
$\lambda \neq \mu$. Thus we find that, similar to the case of NLS
 model [6], the correspondence between Poisson brackets and commutators 
among some elements of monodromy matrix 
may turn out to be a quite nontrivial one. 

Due to eqn.(4.12a) it follows that all operator 
valued coefficients occurring in the expansion of 
 $\ln A(\lambda)$ in powers of $\lambda $
would commute among themselves and, as a consequence, the  monodromy matrix 
(\ref {d1}) of DNLS model leads to a quantum integrable system.
With the help of eqn.(\ref {d1}), it is easy to find that
 $ A(\lambda) \v 0 \r = \v 0 \r$. By using this relation
and eqn.(4.12c),  it can be shown that
\beq
 A(\lambda ) \, \v \mu_1 ,\mu_2  , \cdots , \mu_N  \r ~=~
\prod_{r=1}^N  \left( { \mu_r^2 q - \lambda^2 q^{-1} 
\over \mu_r^2  - \lambda^2 -i \epsilon } \right)  \, 
  \v \mu_1 ,\mu_2  , \cdots , \mu_N  \r \, ,
\label {d13}
\eeq
where $\mu_j$s are all distinct real numbers and $
  \v \mu_1 ,\mu_2  , \cdots , \mu_N  \r \equiv 
 B^\dagger(\mu_1) B^\dagger(\mu_2) \cdots B^\dagger(\mu_N) \v 0 \r $.
Thus the states 
  $\v \mu_1 ,\mu_2  , \cdots , \mu_N  \r $ diagonalise the generator of 
conserved quantities for the quantum
 DNLS model. However, by using eqn.(\ref {d13}), one finds that  the 
eigenvalues corresponding to different expansion coefficients of 
 $ \ln A(\lambda) $ would be complex quantities in general.
To extract real eigenvalues, we define another operator 
$ {\hat A}(\lambda)$ through the relation: 
$  {\hat A}(\lambda) \equiv 
  A(\lambda  e^{ -{i \alpha\over 2}}) $ and expand $\ln {\hat A}(\lambda)$ as
\beq
 \ln {\hat A}(\lambda) = 
\sum_{n=0}^{\infty} \frac{ i \,{\cal C}_n}{\lambda^{2n}}  \, .
\label {d14}
\eeq
With the help of eqns.(\ref {d13}) and (\ref {d14}) it is easy to see that
${\cal C}_n$s satisfy eigenvalue equations like 
${\cal C}_n \,
  \v \mu_1 ,\mu_2  , \cdots , \mu_N  \r = \kappa_n \,
  \v \mu_1 ,\mu_2  , \cdots , \mu_N  \r $,
 where the first few $\kappa_n$s are explicitly given by 
\beq 
\kappa_0 =  \alpha N, ~~ \kappa_1 = 2 \sin \alpha \sum_{j=1}^N \mu_j^2 , ~~
\kappa_2 =  \sin 2\alpha \sum_{j=1}^N \mu_j^4 \, .
\label {d15}
\eeq
In analogy with the classical case, one may now
 define the Hamiltonian for quantum DNLS model as
$ {\cal H} = - \frac{1}{8 \xi} {\cal C}_2 $. 
Eigenvalue equations corresponding to this Hamiltonian are evidently
 given by
\beq
 {\cal H} \v \mu_1 ,\mu_2  , \cdots , \mu_N  \r 
 ~=~  + \frac{1}{4} \sqrt{ 1- \xi^2}
 \left( \sum_{j=1}^N \mu_j^4 \right) \, 
  \v \mu_1 ,\mu_2  , \cdots , \mu_N  \r \, .
\label {d16}
\eeq

Till now we have assumed that $\mu_j$s are some real parameters, for which 
 $ \v \mu_1 ,\mu_2  , \cdots , \mu_N  \r $ represents a scattering state.
One can also construct the quantum $N$-soliton state for DNLS model [12]
by choosing complex values of $\mu_j$ given by 
\beq
  \mu_j ~=~ \mu \, \exp \left[ - i \alpha \left( {N+1 \over 2} -j \right)
\right] \, ,
\label {d17}
\eeq
where $\mu $ is a real parameter and $j\in [1,2, \cdots N]$. 
Thus $\mu_j$s are uniformly distributed on a circle of radius $\mu$.
For this choice of 
 $\mu_j$, eqn.(\ref {d13}) reduces to a simple form like 
\beq
 A(\lambda ) \, \v \mu_1 ,\mu_2  , \cdots , \mu_N  \r ~=~
q^{-N}  \left(  { \lambda^2  - \mu^2 q^{N+1} \over 
  \lambda^2  - \mu^2 q^{-N+1}  } \right)
  \v \mu_1 ,\mu_2  , \cdots , \mu_N  \r \, ,
\label {d18}
\eeq
where the eigenvalue
of $A(\lambda)$ has only one zero and one pole 
on the complex $\lambda^2$-plane.  
By using (\ref {d18}), we obtain the energy eigenvalue
corresponding to the quantum $N$-soliton state as
\beq
 {\cal H} \v \mu_1 ,\mu_2  , \cdots , \mu_N  \r 
 ~=~   { \mu^4 \sin (2\alpha N) \over 8 \sin \alpha }
  \v \mu_1 ,\mu_2  , \cdots , \mu_N  \r \, .
\label {d19}
\eeq
In general, we may choose 
any positive integer value of $N$ (greater than one)
for constructing a quantum soliton state. However, we now consider 
the DNLS model with 
 some particular values of coupling constant
given by $\xi = - \sin \alpha = -\sin \left({2 \pi m \over n}
\right) $, where $m$ and $n$ are 
 nonzero integers which do not have any common factor. By using 
eqn.(\ref {d17}),  one obtains 
 $\mu_j = \mu_{j+n}$ for this case. Since all $\mu_j$ must take distinct
values, we get $N \leq n$ as a restriction on the number of quasi-particles 
which form a bound state for the quantum DNLS model corresponding 
to coupling constant $\xi = - \sin \left ( {2\pi m \over n} \right) $.

Thus, by applying the method of algebraic Bethe ansatz, we are able to 
construct the exact eigenstates for quantum DNLS model.  
The commutation relation (4.12e)  also 
plays an important role in the framework
of algebraic Bethe ansatz, since by using this commutation relation
one should be able to calculate the norm of eigenstates 
  $\v \mu_1 ,\mu_2  , \cdots , \mu_N  \r $ and 
various correlation functions of the DNLS system. However,
it may be noted that the commutation relation 
(4.12e) contains product of generalised functions $( \lambda^2
- \mu^2 - i \epsilon )^{-1}( \lambda^2 - \mu^2 + i \epsilon )^{-1} $, 
which does not make sense at the limit $\lambda 
\rightarrow \mu$.  As a result, the action of operators 
 $B^\dagger(\lambda),  B(\mu) $ are not well defined on the Hilbert 
space [8] and eigenstates like 
  $\v \mu_1 ,\mu_2  , \cdots , \mu_N  \r $ 
are not normalised on the $\delta$-function. However, it is well known that,
one can avoid this type of problem in the case of NLS
equation by considering the quantum analogue of classical reflection
operators [1,6,7]. So, in analogy with the case of 
 NLS equation, at present we consider a reflection
 operator given by
\beq 
R^{\dagger}(\lambda ) ~=~  A^{-1}(\lambda)B(\lambda) \, ,
\label{d20}
\eeq
and its adjoint $R(\lambda ) $. By using eqns.(4.12a-e), 
we find that such reflection operators satisfy well defined 
commutation relations like 
\bea
&&R(\lambda)R(\mu) = S^{-1}(\lambda , \mu) \, R(\mu) R(\lambda) \, , \nn \\
&&R^{\dagger}(\lambda)R^{\dagger}(\mu) 
= S^{-1}(\lambda , \mu) \, R^{\dagger}(\mu) R^{\dagger}(\lambda) \, , \nn \\
&&R(\lambda)R^{\dagger}(\mu) 
= S(\lambda , \mu) \, R^{\dagger}(\mu) R(\lambda) + 
4 \pi \lambda^2 \delta(\lambda^2 - \mu^2) \, ,
\label {d21}
\eea
where 
\beq
S(\lambda , \mu) \, = \,  { {\lambda^2 q - \mu^2 q^{-1}} \over 
{\lambda^2 q^{-1} - \mu^2 q }} \,. 
\label{d22}
\eeq
It is evident that these commutation relations  among 
reflection operators of DNLS model are 
nicely encoded in a form of Zamolodchikov-Faddeev algebra [1,13] and 
 $S(\lambda , \mu) $ (\ref {d22}) represents the nontrivial
 $S$-matrix element of two-body scattering between the 
related quasi-particles.  It is easy to check that this 
$S(\lambda , \mu) $ satisfies the conditions given by
\beq 
S^{-1}(\lambda , \mu) = S(\mu ,\lambda) = S^*(\lambda , \mu) \, ,
\label {d23}
\eeq
and remains nonsingular at the limit $\lambda \rightarrow \mu$.  
As a result,  the action of operators 
 $R^\dagger(\lambda),  R(\mu) $ would be well defined on the Hilbert 
space and eigenstates like 
 $R(\mu_1) R(\mu_2) 
\hfil \break  
\cdots R(\mu_N) \v 0 \r $ 
can be normalised on the $\delta$-function. 

\medskip

\noindent \section {Concluding Remarks}
 
We find that the classical monodromy matrix for DNLS model 
can be written as a $SU(2)$ ($SU(1,1)$)  group valued 
object for positive (negative) value of the corresponding coupling
constant. By using such symmetric form of classical monodromy matrix, 
we derive Poisson bracket
 relations among the scattering data of the DNLS model
for all values of the coupling constant.
We also quantise the monodromy matrix of DNLS model on a finite 
interval. A variant of quantum inverse scattering method, which can be 
applied to field models without performing any lattice regularisation,
fixes all parameters in the quantum monodromy matrix of DNLS model 
in a nontrivial way. Similar to the classical case, this quantum monodromy 
matrix exhibits $U(2)$ ($U(1,1)$)  symmetry 
for positive (negative) value of the coupling constant. By applying 
quantum inverse scattering method, we derive all possible 
commutation relations among the elements of such monodromy matrix. 
 Infinite interval limits of these commutation relations enable us to 
  construct the exact eigenstates for quantum DNLS model
 through algebraic Bethe ansatz.  In this context, we consider 
the DNLS model with 
 some special values of coupling constant
given by $\xi = - \sin \alpha = -\sin \left({2 \pi m \over n}
\right) $, where $m$ and $n$ are 
 nonzero integers which do not have any common factor. 
It turns out that the number of quasi-particles, 
 which form a bound state for such quantum DNLS model, can not exceed the
value $n$.  

We also obtain the commutation relation between creation and annihilation 
operators associated with quasi-particles of DNLS model
and find out the $S$-matrix for two-body scattering.
Such a commutation relation between 
creation and annihilation operators 
should play an important role in a future study,
 since by using it 
one might be able to calculate the norm of Bethe eigenstates and 
various correlation functions of the DNLS system. 
It may be noted that, there exist quantum integrable multicomponent 
 generalisations of NLS model which can be diagonalised 
through algebraic Bethe ansatz [14-17].
  A large class of multicomponent 
 classical DNLS models, having infinite number of conserved quantities, 
are also studied in the literature
[18,19].  However, the Hamiltonian structure of such 
multicomponent DNLS models have not yet received much attention.
So it might be interesting to investigate whether there exist 
  some multicomponent generalisations of
 classical DNLS model which are integrable in the 
Liouville sense and retain their integrability property even after 
quantisation.

\medskip


\newpage

\leftline {\large \bf Appendix}

\baselineskip=14pt
\medskip

A direct attempt to calculate ${ \d \over \d x_2} \left( 
{\cal T}^{x_2}_{x_1}(\lambda)\otimes {\cal T}^{x_2}_{x_1}(\mu) \right)$ 
 by using eqn.(\ref {c4}) evidently leads to indeterminate 
expressions of the form $\left[ {\cal T}^{x_2}_{x_1}
(\lambda), \psi^\dagger( x_2 ) \right]$. To bypass this problem, we follow 
the method of extension  [2] which shifts the upper limit of one monodromy 
matrix (say ${\cal T}^{x_2}_{x_1}(\lambda) $)
by introducing a small parameter $\epsilon $ and takes  $\epsilon 
\rightarrow 0$ limit only after differentiating the product 
${\cal T}^{x_2 +\epsilon }_{x_1}(\lambda)\otimes {\cal T}^{x_2}_{x_1}(\mu)$
with respect to $x_2$. Thus, by using eqn.(\ref {c4}),  we obtain
\bea
\frac{\partial}{\partial x_2} 
\left ( \, {\cal T}^{x_2 + \epsilon}_{x_1}( \lambda ) \otimes 
{\cal T}^{x_2}_{x_1}( \mu ) \, \right)
 &=&   \vdots \,  \left( \, {\cal U}_q( x_2 + \epsilon , \lambda )
\otimes \one + \one \otimes {\cal U}_q( x_2 , \mu ) \, \right)
{\cal T}^{x_2 + \epsilon}_{x_1}( \lambda ) \otimes 
{\cal T}^{x_2}_{x_1}( \mu ) \, \vdots  \nn \\
&~&+ \,  K_+ \, + \,  K_- ~ ,  
~~~~~~~~~~~~~~~~~~~~~~~~~~~~~~~~~~~~~~~~~~~~~~~~~(A1) \nn
\eea 
where
\bea
K_+ &=& i \xi \mu \left[ {\cal T}^{x_2 + \epsilon}_{x_1}( \lambda ),
\psi^\dagger( x_2 ) \right] \otimes \sigma_+ {\cal T}^{x_2}_{x_1}( \mu ) +
i f \left[ {\cal T}^{x_2 + \epsilon}_{x_1}( \lambda ),\psi^\dagger( x_2 ) 
\right] \otimes 
\nn \\ 
&~&~~~~~~~~~~ e_{11} {\cal T}^{x_2}_{x_1}( \mu ) \psi( x_2 ) 
- i g \left[ {\cal T}^{x_2 + \epsilon}_{x_1}( \lambda ),\psi^\dagger( x_2 ) 
\right] \otimes e_{22} {\cal T}^{x_2}_{x_1}( \mu ) \psi(x_2) \, , 
\nn \\
K_- &=& i \lambda \sigma_- {\cal T}^{x_2 + \epsilon}_{x_1}( \lambda ) \otimes 
\left[ \psi( x_2 + \epsilon ), {\cal T}^{x_2}_{x_1}( \mu ) \right] 
+ i f \psi^\dagger( x_2 + \epsilon )e_{11} 
{\cal T}^{x_2 + \epsilon}_{x_1}( \lambda ) \otimes 
\nn \\
&~& ~~~~\left[ \psi( x_2 + \epsilon ), {\cal T}^{x_2}_{x_1}( \mu ) \right] 
- i g \psi^\dagger( x_2 + \epsilon ) e_{22}{\cal T}^{x_2 + \epsilon}_{x_1}
( \lambda ) \otimes \left[ \psi( x_2 + \epsilon ), 
{\cal T}^{x_2}_{x_1}( \mu ) \right]  \, .\nn
\eea

Now we consider the case $\epsilon > 0$. 
Since $\psi( x_2 + \epsilon)$ commutes with 
$\psi( x ), \, \psi^\dagger( x )$ for all $x$
lying within $x_1$ and $x_2$,  we can write
$\left[ \psi( x_2 + \epsilon ), {\cal T}^{x_2}_{x_1}( \mu ) \right] =0$ and 
 $K_- = 0 $  for this case. So, we have to calculate only the
nontrivial commutator $\left[ {\cal T}^{x_2 + \epsilon}_{x_1}
( \lambda ), \psi^\dagger( x_2 ) \right]$ which appears in the expression 
of $K_+$.  To this end, we consider a `transformation' $\Omega$ 
which replaces the classical variables $\psi(x)$ and $\psi^*(x)$  
by quantum operators $\psi(x)$ and $\psi^\dagger(x)$ respectively 
($\Omega^{-1}$ denotes the reverse transformation). By applying a
 correspondence principle [2] to the present case, we obtain 
\bea
~~~~~~~~~~~~~~\left[ {\cal T}^{x_2 + \epsilon}_{x_1}( \lambda ), 
\psi^\dagger( x_2 ) \right]
= i  : \, \Omega  \,  \left \{ T^{x_2 + \epsilon}_{x_1}( q; \lambda ), 
\psi^*( x_2 ) \right \} \, : \, , \nn 
~~~~~~~~~~~~~~~~~~~~~~~~~~~~~ (A2)
\eea 
where
$ T^{x_2 + \epsilon}_{x_1}( q; \lambda )$  represents a classical 
monodromy matrix given by
$$
T^{x_2 + \epsilon}_{x_1}( q; \lambda ) =
\quad  {\cal P} \exp \int_{x_1}^{x_2 + \epsilon } 
U_q(x,\lambda) dx   \, ,
$$
and $U_q( x, \lambda ) = \Omega^{-1} {\cal U}_q( x , \lambda ) $.
 By using the fundamental PB relations
(\ref {a4}), it is easy to find that
\bea
\{ T^{x_2 + \epsilon}_{x_1}( q; \lambda ), \psi^*( x_2 ) \} &=& 
\int^{x_2 + \epsilon}_{x_1} dx \,  T^{x_2 + \epsilon}_x( q; \lambda ) 
\left \{ U_q( x, \lambda ), \psi^*( x_2 ) 
\right \} T^x_{x_1}( q; \lambda ) \nn \\
&=& T^{x_2 + \epsilon}_{x_2}( q; \lambda ) \left( f \psi^*( x_2 ) e_{11} - g
\psi^*( x_2 ) e_{22} + \lambda \sigma_- \right) 
T^{x_2}_{x_1}( q; \lambda ) \, . \nn
\eea  
Taking  $\epsilon \rightarrow 0$ limit of the above expression and 
inserting it to (A2),  we obtain
\bea
~~~~~~\lim_{\epsilon \rightarrow 0} \,
\left[ \, {\cal T}^{x_2 + \epsilon}_{x_1}( \lambda )
, \psi^\dagger( x_2 ) \, \right] = i \left( f \psi^\dagger( x_2 ) e_{11} - g
\psi^\dagger( x_2 ) e_{22} + \lambda \sigma_- \right) \, {\cal T}^{x_2}_{x_1}
(\lambda) \, . \nn 
~~~~~~~~~~~~~  (A3)
\eea
Taking $\epsilon \rightarrow 0$ limit also in eqn.(A1) and 
using (A3), we finally obtain the differential 
equation (\ref{c5}). Note that, instead of $\epsilon >0$,
 we could have chosen $ \epsilon < 0 $ in eqn.(A1).  Only the commutator 
$\left[ \psi( x_2 + \epsilon ), {\cal T}^{x_2}_{x_1}( \mu ) \right]$ gives 
a nontrivial contribution for this case.
However, repeating similar steps as outlined above and finally taking 
the $\epsilon \rightarrow 0$ limit, 
we get again the same differential equation (\ref{c5}). 

\newpage

\leftline {\large \bf References}
\medskip
\begin{enumerate}
\item L.D. Faddeev, Sov. Sci. Rev. C1 (1980) 107; in {\it Recent Advances in
Field Theory and Statistical Mechanics, } ed. J.B.Zuber and R.Stora ,
 (North-Holland, Amsterdam, 1984 ) p.561. 

\item  E.K. Skylanin, in  
Yang-Baxter Equation in Integrable systems, Advanced series in Math. Phys. 
Vol. 10, edited by M. Jimbo ( World Scientific, Singapore, 1990) p.121.

\item V. E. Korepin, N. M. Bogoliubov, and A. G. Izergin, {\it Quantum Inverse 
Scattering Method and Correlation Functions} (Cambridge Univ. Press, 
Cambridge, 1993) and references therein. 

\item  Z.N.C. Ha, {\it Quantum Many-Body Systems in One Dimension} 
(World Scientific, Singapore, 1996) and references therein.

\item J. Honerkamp, P. Weber, A. Wiesler, Nucl. Phys. B 152 (1979) 266.

\item 
 H.B. Thacker and D. Wilkinson, Phys. Rev. D 19 (1979) 3660;
D.B. Creamer, H.B. Thacker and D. Wilkinson, Phys. Rev. D 21 (1980) 1523.

\item K. M. Case, J. Math. Phys. 25 (1984) 2306.

\item E. Gutkin, Phys. Rep. 167 (1988) 1.

\item D.J. Kaup and A.C. Newell, J. Math. Phys. 19 (1978) 798.

\item A. Kundu, J. Phys. A 21 (1988) 945.

\item H.H. Chen, Y.C. Lee and C.S. Liu, Phys. Scr. 20 (1979) 490.

\item A. Kundu and B. Basu-Mallick, J. Math. Phys. 34 (1993) 1052.

\item A.B. Zamolodchikov and A.B. Zamolodchikov, Ann. Phys. 120 (1979) 253. 

\item F.C. Pu and B.H. Zhao, Phys. Rev. D 30 (1984) 2253.

\item F.C. Pu and B.H. Zhao, Nucl. Phys. B 275 [FS 17] (1986) 77.

\item S. Murakami and M. Wadati, J. Phys. A 29 (1996) 7903.

\item M. Mintchev, E. Ragoucy, P. Sorba and P. Zaugg, 
 J. Phys. A 32 (1999) 5885.

\item T. Tsuchida and M. Wadati, Phys. Lett. A 257 (1999) 53; Inv. Prob. 
15 (1999) 1363. 

\item P.J. Olver and V.V. Sokolov, Inv. Prob. 14 (1998) L5.

\end{enumerate}

\end{document}